\documentclass[usenatbib]{mnras}

\usepackage{newtxtext,newtxmath}

\usepackage[T1]{fontenc}

\usepackage{url}
\usepackage{longtable}
\usepackage{lscape}
\usepackage{multirow}
\usepackage{color}
\usepackage{textcomp}
\usepackage{natbib}
\usepackage{footnote}
\usepackage{graphicx}	
\usepackage{amsmath}	
 
\usepackage{amssymb}

\title[The distances of 61 PGCCs]{The distances of 61 PGCCs in the Second Galactic Quadrant}

\author[Guo et al.]{
H.-L. Guo,$^{1}$ 
B.-Q. Chen,$^{1}$\thanks{E-mail: bchen$@$ynu.edu.cn (BQC); gxli$@$ynu.edu.cn (GXL); x.liu$@$ynu.edu.cn (XWL) }
 G.-X. Li,$^{1}$\footnotemark[1]
 Y. Huang, $^{1}$ 
 Y. Yang, $^{1}$ X.-Y. Li, $^{1}$
 W.-X. Sun, $^{1}$
and X.-W. Liu$^{1}$\footnotemark[1]
\\
$^{1}$South-Western Institute for Astronomy Research, Yunnan University, Chenggong District, Kunming 650091, P.\,R. China\\
}

\date{Accepted XXX. Received YYY; in original form ZZZ}

\pubyear{2020}

\begin{document}
\label{firstpage}
\pagerange{\pageref{firstpage}--\pageref{lastpage}}
\maketitle

\begin{abstract}
Determining the distances to the Planck Galactic cold clumps (PGCCs) is crucial for the measurement of their physical parameters and the study of their Galactic distribution. Based on two large catalogues of stars with robust distances and reddening estimates from the literature, we have estimated accurate distances to 61 PGCCs in the second Galactic quadrant. For this purpose, we have selected stars along the sightlines overlapping with the cores of the sample clumps and fitted the reddening profiles with a simple reddening model. The typical uncertainties of the resultant distances of these PGCCs are less than 8 per\,cent.  The new estimates differ significantly from the kinematic values, well known to suffer from large errors. With the new distances, we have updated the physical properties including the radii, masses and virial parameters of the cores of the PGCCs.
\end{abstract}


\section{Introduction}
Star formation is a key process in the life cycle of galaxies. It occurs in the cold and dense parts of molecular clouds \citep{Blitz1999}, whose physical and chemical properties are still poorly understood. A statistical study of the properties of the cold dense clumps within  molecular clouds based on unbiased large surveys of the Milky Way is important to understand the initial conditions and the early phases of star formation.

The {\it Planck} mission \citep{Planck2011a} has revealed a huge number of Planck Galactic cold clumps (PGCCs) based on multi-band observations ranging from sub-millimetre to millimetre wavelengths. The Cold Core Catalogue of Planck Objects \citep[C3PO;][]{Planck2011d} containing $\sim$ $10^4$ cold cores, and the sub-catalogue, the Planck Early Release Cold Cores Catalogue \citep[ECC;][]{Planck2011b} containing 915 most reliable detections, were released in 2011. The C3PO presented the first unbiased, all-sky catalogue of cold objects. PGCCs are cold (10-15\,K) and turbulence-dominated. They have relatively low column densities compared to other star forming regions \citep{Planck2011c, Planck2011d, Wu2012}. PGCCs are widely used to probe the initial conditions of star formation and the early evolution of stars in a wide range of Galactic environments. Understanding the properties of those clumps also offer a unique view of the evolution of molecular gas.

Accurate estimation of distances to PGCCs is crucial for both the determinations of their physical properties, as well as their Galactic distribution. In most studies, the commonly adopted distances are the kinematic values yielded with a Bayesian estimator \citep{Reid2016}, based on the locations on the celestial sphere and radial velocities of the sources, with knowledge such as the Galactic rotation curve \citep[e.g.][]{Reid2014, Xue2015, Huang2016} and the distances to the previously known star-forming regions \citep[e.g.][]{Pestalozzi2005, Xu2013, Zhang2013, Choi2014, Wu2014, Hachisuka2015, Xu2016} taken as priors. Although the approach has been widely used to determine the distances of star forming regions in the Galaxy, the method is not particularly suitable for PGCC. This is because PGCCs are mostly nearby clouds of relatively small masses \citep{Zhang2018} and their motion might not follow that of the Galactic rotation exactly  \citep{Kolpak2003, Reid2009}. Consequently, for PGCCs, their kinematic distances often suffer from large errors, leading to significant difficulties for the subsequent analyses.

Direct estimates of distances to PGCCs can be achieved via the
three-dimensional (3D) dust extinction mapping. Compared to the diffuse interstellar medium, PGCCs have much higher densities of dust grains. The amount of sightline dust extinction is thus expected to increase significantly and abruptly at the location
of a PGCC. One can thus estimate the distance to a PGCC by finding the position of a sharp increase of extinction in the extinction versus distance profile of the sightline towards the PGCC. Benefiting from accurate extinction and distance estimates of millions of individual stars\citep[e.g.,][]{Chen2014, Chen2019, Green2015, Green2018, Green2019, Lallement2019, Anders2019, Guo2020}, accurate distances to large numbers of Galactic molecular clouds have been obtained in this way \citep[e.g.,][]{Schlafly2014, Chen2017b, Chen2020b, Chen2020}, as well as distances to other types of objects, including supernovae remnants associated with molecular clouds \citep[e.g.,][]{Zhao2018, Yu2019, Shan2019, Zhao2020, Wang2020}. The robustness of distances thus obtained relies on the precision of distance estimates of the individual stars. Tests based on stellar distances yielded by Gaia DR2 parallaxes \citep{Gaia2016, Gaia2018a, Lindegren2018} have shown that molecular cloud distances estimated with the aforementioned extinction method can reach an accuracy better than $\sim$ 5 per\,cent \citep{Chen2020}.

In the current work, we have applied the method and derived accurate distances to a sample of 64 PGCCs, of which \citet{Zhang2018} presented a detailed study of their physical properties based on the kinematic distances calculated with the Bayesian distance calculator of \citet{Reid2016}. For this purpose, we have used the stellar catalogues from \citet{Chen2019} and \citet{Green2019}. Apart from providing significantly improved distance estimates, our results also allow us to assess the uncertainties of kinematic distances and provide an update of the physical parameters presented by \citet{Zhang2018}.

\section{Data}
\label{data}
\citet{Zhang2016} selected 96 PGCCs in the second Galactic
quadrant (98\degr $< l <$ 180\degr and $-$4\degr $< b <$ 10\degr) which are those densest ECCs in the region. \citet{Zhang2018} studied 64 of the 96 \citet{Zhang2016} PGCCs that were covered by both the Submillimetre Common-User Bolometer Array 2 \citep[SCUBA-2;][]{Holland2013} 850\,${\mu}$m continuum and Purple Mountain Observatory (PMO) $^{13}$CO and C$^{18}$O $J$= 1-0 line observations. These sources are of great important to understand the early evolution of molecular clouds and dense cores in different environments. To estimate the extinction distances of the PGCCs catalogued by \citet{Zhang2018}, one needs extinction profiles along the sightlines of those PGCCs. In the current work, data from two previous studies have been used to construct those profiles.

\citet{Chen2019} presented estimates of dust reddening, and colour excesses $E(G - K_{\rm S} )$, $E(G_{\rm BP} - G_{\rm RP})$ and $E(H - K_{\rm S})$ of over 56 million stars in the Galactic disk ($|b|<$ 10\degr), based on the optical to near-infrared (IR) photometry provided by the Gaia DR2, the Two Micron All Sky Survey \citep[2MASS;][]{Skrutskie2006} and the Wide-Field Infrared Survey Explorer \citep[WISE;][]{Wright2010}. The reddening values were estimated with a Random Forest model trained by a training sample constructed from the various spectroscopic surveys. In this work, we convert their $E(G_{\rm BP} - G_{\rm RP})$ values to those of  $E(B - V)$ using the relation, $E(B - V)$ = 0.75$E(G_{\rm BP} - G_{\rm RP})$ \citep{Chen2019}. The typical uncertainties in $E(B-V)$ are $\sim$ 0.07\,mag \citep{Chen2019}. Distances from \citet{Bailer2018} are adopted for the stars, estimated from the Gaia DR2 parallaxes with a simple Bayesian approach. \citet{Chen2019} excluded the stars with Gaia DR2 parallax uncertainties larger than 20 per\,cent, leading to a sample of over 32 million stars. The typical distances probed by stars catalogued by Chen et al. are only about 4\,kpc, mainly due to their parallax error cut adopted in that work. 

In this work, we have also used the extinction catalogue presented by \citet{Green2019}. Based on optical to near-IR photometry provided by the Pan-STARRS1 \citep{Chambers2016}, 2MASS and Gaia DR2, as well as the parallaxes from the Gaia DR2, \citet{Green2019} have calculated distances and reddening values of about 799 million stars  in the northern sky (Declination $\delta > -$30\degr) with a hierarchical Bayesian algorithm. The stellar parameter catalogue of \citet{Green2019} provides values of distance modulus (``dm") and reddening (``E") of the stars. The typical uncertainties are $\sim$ 0.10\,mag for $E(B - V)$ \citep{Green2019}. The typical depth distance limit of sightlines that overlap with our sample
PGCCs are about 7.0\,kpc.

\begin{figure*}
\centering
\includegraphics[width=0.48\textwidth, angle=0]{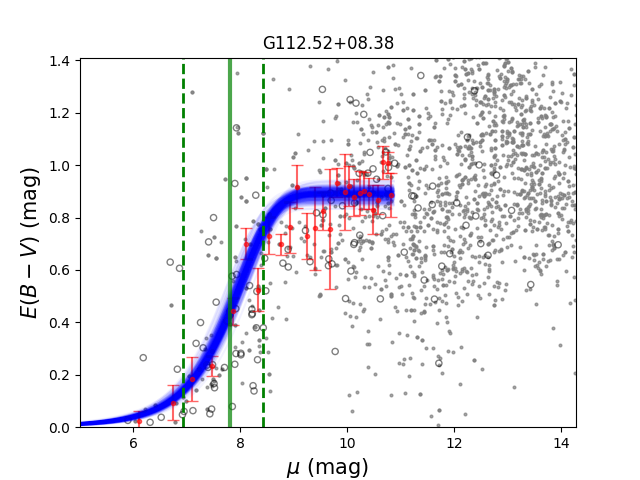}
\includegraphics[width=0.48\textwidth, angle=0]{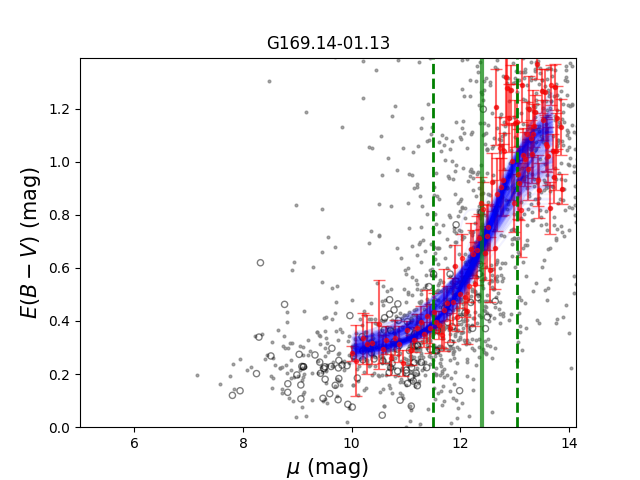}
\includegraphics[width=0.48\textwidth, angle=0]{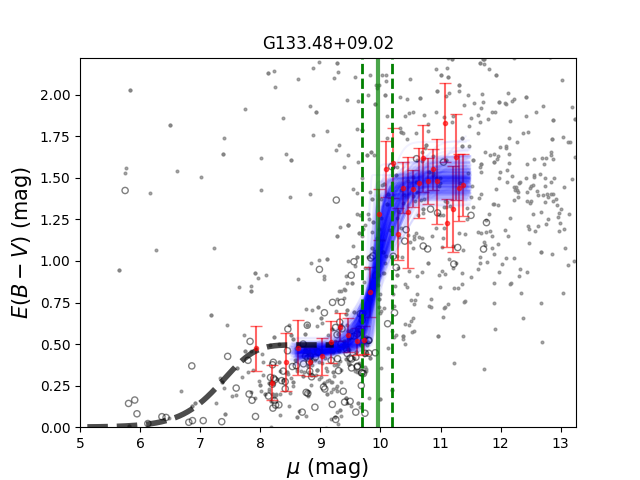}
\includegraphics[width=0.48\textwidth, angle=0]{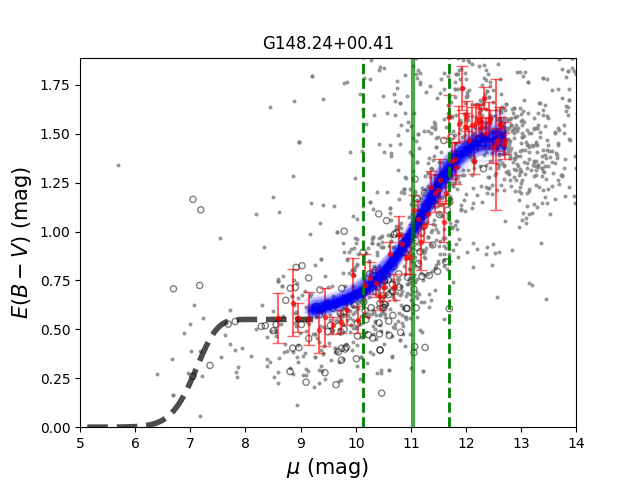}

\caption{Distance determinations for four example clumps G112.52+08.38 (upper left), G169.14-01.13 (upper right), G133.48+09.02 (lower left) and G148.24+00.41 (lower right). X-axis denote the distance modulus while y-axis the colour excesses. In each panel, the grey unfilled circles and grey dots represent individual stars selected respectively from \citet{Chen2019} and \citet{Green2019}. Red dots and error bars are respectively median values and standard errors of $E(B-V)$ in the individual distance bins of bin-size 50\,pc. The blue lines are the best-fit reddening laws based on the 300 randomly generated samples of stars. The vertical green solid and dashed lines mark respectively the best-fit distances ($d_{0}$) and the widths ($\delta d$) of the PGCCs. The black dashed lines in the bottom two panels show two additional extinction jumps.}
\label{clumps}
\end{figure*}

\begin{figure}
\centering
\includegraphics[width=0.45\textwidth, angle=0]{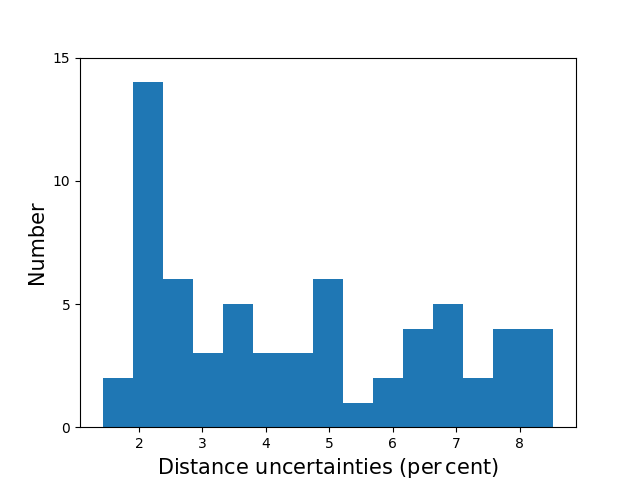}
\caption{Distribution of the resultant distance uncertainties of the catalogued PGCCs.}
\label{errdis}
\end{figure}

\section{Method}
\label{method}

\begin{figure*}
\centering
\includegraphics[width=0.48\textwidth, angle=0]{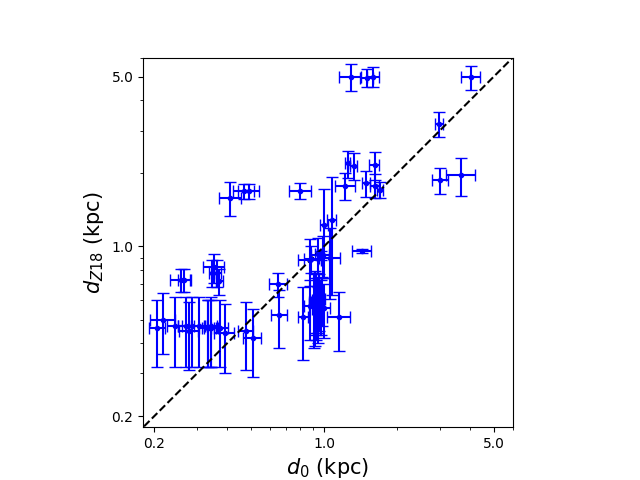}
\includegraphics[width=0.48\textwidth, angle=0]{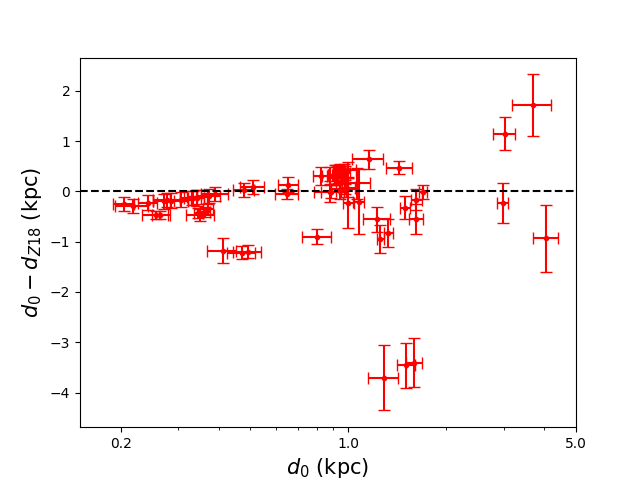}

\caption{Comparison of our extinction distances with their kinematic values from \citet{Zhang2018}. In the left panel, the black dash line denoting complete equality is used to guide the eyes.}
\label{compare}
\end{figure*}

\begin{figure}
\centering
\includegraphics[width=0.45\textwidth, angle=0]{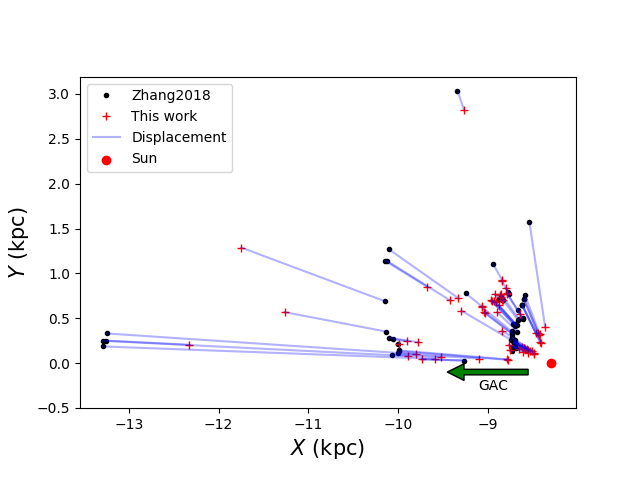}
\caption{The spatial distributions of the catalogued PGCCs based on the extinction distances presented in the current work (red pluses) and on the kinematic distances from \citet{Zhang2018} (black dots).
The Sun, located at $(X, Y)$ = ($-$8.3, 0.0)\,kpc, is
marked by the red filled circle in the Figure. The green arrow indicates the direction of the Galactic anti-centre.}
\label{xy}
\end{figure}

To determine the extinction distance to a given sample PGCC, we first select stars from the catalogues of \citet{Chen2019} and \citet{Green2019} along the lines of sight overlap with the PGCC. For each PGCC, there are several cores identified by \citet{Zhang2018}. In the current work, we only use the reddening profiles of the sightlines overlapping with those cores, where the sharp increases of the reddening values, i.e. the reddening jumps, are much more significant than for the other regions of the cloud. For each core of a given PGCC, we select stars that fall within the core size area \citep[`FWHM' in Table 3 of][]{Zhang2018}. For a  typical solid angle of about 6.15\,arcmin${^2}$ subtended by a core, there are typically $\sim$ 300 stars within the area. In the current work, we assume that all the cores
in a single clump have the same distances and merge all the stars selected from the individual cores of a clump to determine its extinction distance. Typically, there are $\sim$ 1,900 stars for an area of about 61.2\,arcmin${^2}$ covered by a typical clump. There are not enough stars ($N <$ 1000) for 6 catalogued clumps (G133.28+08.81, G133.48+09.02, G142.62+07.29, G147.01+03.39, G156.04+06.03 and G176.35+01.92), mainly due to the small total areas of their cores. This leads to very large distance uncertainties of those clumps. Accordingly, to better constrain the distances of those six clumps, we have adopted a star selection radius of twice the core size for each core. For the nearby clumps (distance $d <$ 0.5\,kpc), we are not able to find enough foreground stars to the clumps within the radii of their cores. For sightlines through those nearby clumps, we also adopt a larger star selection radius twice of their core sizes.

We then find the position of the reddening jump, i.e. the
distance of the PGCC, by modeling the reddening profile constructed from the selected stars. As the dust grain densities in the cores of a PGCC are much higher than those in the diffuse medium, we have assumed that the dust extinction within the distance range of the clump all arises from the local dust grains within the cores of the clump. The reddening profile $E(d)$ within the distance range of the clump is thus given by,
\begin{equation}
E(d) = E^0 + E^1(d),
\end{equation}
where $E^0$ is a constant that represents the reddening contributed by the foreground dust, and $E^1(d)$ is the reddening contributed by the dust grains in the cores of the clump. Assuming a Gaussian radial density profile of the core dust, we have,
\begin{equation}
E^1(d) = \frac{\delta E}{2}[1 + {\rm{erf}}(\frac{d-d_0}{\sqrt{2}\delta d})],
\end{equation}
where $d_{\rm 0}$ is the distance of the clump, $\delta d$ the width of the reddening jump and $\delta E$ the total reddening contributed by the cores of the clump.

There could be more than one reddening jump in the reddening profile of a given PGCC. We expect that the cores of the PGCC would produce the most significant jump. We thus only fit the most significant jump for a given profile. The binned average reddening values in the distance range of a jump are fitted with the above model using the Markov chain Monte Carlo (MCMC) 
algorithm {\sc emcee} \citep{Foreman2013}. The uncertainties of the resultant parameters are calculated using the Monte Carlo method \citep{Jenkins2003}. For each clump, we randomly generate 300 samples of the selected stars according to their distance and reddening uncertainties. The fitting algorithm is applied to all those samples. The rms scatters of the derived parameters are taken as their uncertainties.

\section{Result}

The distance estimation algorithm outlined above is applied to all the 64 PGCCs catalogued. Among these, three clumps (G175.20+01.28, G175.53+01.34, G176.35+01.92) exhibit no obvious jump, thus we can not derive their distances. As a result, we have derived directly-measured extinction distances to 61 PGCCs. In Fig.\,\ref{clumps} we show 4 examples of the reddening profile fitting. In general, the extinction profiles yielded by stars from \citet{Chen2019} are in good agreement with those of \citet{Green2019}. Reddening jumps produced by the cores of the PGCCs are significant and clearly visible. The medians binned reddening values are nicely traced by the best-fit models.

The upper two panels of Fig.\,\ref{clumps} show the example extinction distance determinations for clumps G112.52+08.38 and G169.14-01.13. In both cases, one reddening jump produced by the corresponding clump is clearly visible. The resultant distances of
G112.52+08.38 and G169.14-01.13 are respectively 363$\pm13$\,pc and 3,019$\pm239$\,pc. The width of the reddening jump $\delta d$ of the much more distant clump G169.14-01.13 ($\delta d$ = 1017\,pc) is much larger than that of the nearby clump G112.52+08.38 ($\delta d$ = 120\,pc). This is mainly due to the fact that the widths of the reddening jumps are mainly dominated by the distance uncertainties of stars used to construct the reddening profiles \citep{Chen2020}, and in general, the more distant the stars, the larger the uncertainties of their distances.

In the lower two panels of Fig.\,\ref{clumps}, we show the extinction distance determinations for clumps G133.48+09.02 and
G148.24+00.41. More than one reddening jumps are found
in the reddening profiles of these two PGCCs. As mentioned in Sec.\,\ref{method}, we select the most significant reddening jump in such cases. The resultant distances of G133.48+09.02 and G148.24+00.41 are 980$\pm31$\,pc and 1616$\pm78$\,pc, respectively. Again we find that the width of the reddening jump $\delta d$ of the more distant clump G133.48+09.02 ($\delta d$ = 554\,pc) is much larger than that of the nearby clump G148.24+00.41 ($\delta d$ = 112\,pc). 

Figures analogous to Fig.\,\ref{clumps} for all the 61 PGCCs are available online\footnote{\url{http://paperdata.china-vo.org/guo/dust/PGCCs_61.pdf}}. In Table\,1 we list the best-fit values of distance $d_{\rm 0}$ and of the total reddening $\delta E$ of these clumps. For comparison, we also list their values of kinematic distance $d_{\rm Z18}$ derived by  \citet{Zhang2018}. The sample clumps range in distance from $d_{\rm 0}$ $\sim$ 200 to $\sim$ 4,000\,pc. 

In Fig.\,\ref{errdis}, we show a histogram distribution of the resulting distance uncertainties. For our catalogued clumps, the distance uncertainties are smaller than 8 per\,cent. This is benefits from the accurate parallax measurements of stars from Gaia DR2. The uncertainties of the derived distances include contributions from the statistical and systematic errors. The statistical uncertainties are given by the MCMC analysis, and are very small due to the large sample of selected stars. The systematic uncertainties, resulting from the uncertainties of the distances and reddening values of the individual stars, arise mainly from the uncertainties in the Gaia parallaxes, the stellar atmospheric models, and of the extinction law and possible variations of the latter. In this work, most of the sample clumps are located at nearby distances ($d <$ 2\,kpc), where we can have very accurate stellar parallaxes from Gaia DR2. The typical distance errors for the individual stars from the Chen et al. and Green et al. catalogues are smaller than 10 and 15\,per\,cent, respectively.

In Fig.\,\ref{compare}, we compare our newly derived extinction distances with the kinematic values deduced by \citet{Zhang2018}. In general, the newly estimated extinction distances differ significantly from the kinematic ones. Amongst the sample objects, 10 clumps have similar kinematic distances ($d_{\rm Z18} \sim$ 0.46\,kpc), as they all have similar radial velocities ($V_{\rm LSR}$ $\sim -$8.0 or 3.0\,km\,s$^{-1}$). However, their directly measured extinction distances range from 0.21 to 0.51\,kpc. It is quite possible that the peculiar and non-circular motions of the clumps may have significantly affected the calculation of their kinematic distances \citep{Liszt1981, Gomez2006, Reid2009}. For some PGCCs, such as G176.17-02.10 and G177.14-01.21, the differences between the two sets of distances are very large. The kinematic distance probabilities of those clumps derived with the Bayesian Distance Calculator \citep[see Table\,1 of][]{Zhang2018} are very small (less than 0.5), indicating that those kinematic distances are poorly constrained.

\begin{figure*}
\centering
\includegraphics[width=0.45\textwidth, angle=0]{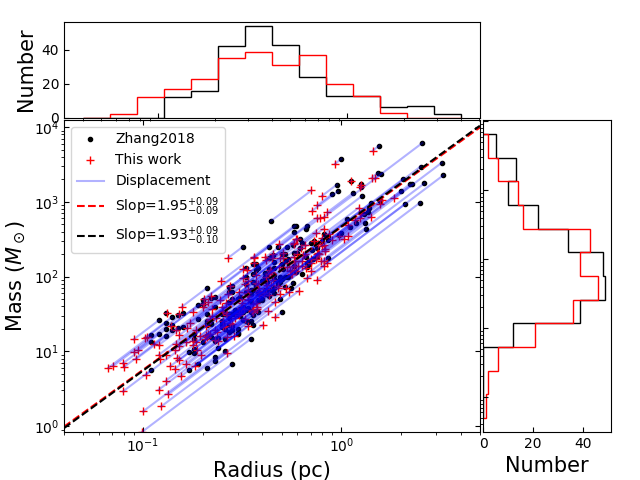}
\includegraphics[width=0.45\textwidth, angle=0]{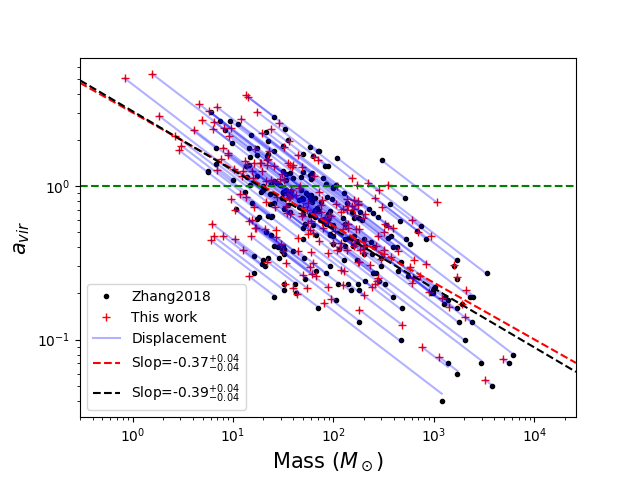}

\caption{{\it Left panel}: Distribution of the individual cores of the catalogued PGCCS in the mass--radius plane. Red pluses and black dots represent respectively the results based on the newly derived extinction distances and on the early kinematic values. Red and black dashed lines give the corresponding best--fit mass--radius relations from the two sets of distances. Histograms of the mass and radius distributions are also plotted on the sides. Red and black histograms represent results based on the extinction distances and on those kinematic ones, respectively. {\it Right panel}: Virial parameter--mass based on our extinction distances (red pluses) and on the previous kinematic distances (black dots). The green dotted line marks the line of $\alpha_{vir}$ = 1.0. The red and black dotted lines are the corresponding best--fit virial parameter-mass relations based on the two sets of distances.}
\label{m_r}
\end{figure*}

\section{Discussion}

The improved distance estimates allow us to revise some of the distance-dependent physical properties of the cores of the PGCCs, such as their radii, masses and the virial parameters. We update those physical parameters by simply rescaling the results of \citet{Zhang2018}. For example, effective radius $R_{\rm eff}$ of a Planck core is proportional to distance $d$, $R_{\rm eff} \propto d$. Thus the revised radius $R'_{\rm eff}$  is evaluated as, 
\begin{equation}
    R'_{\rm eff} = R_{\rm eff} \times d' / d ;,
\end{equation}
Similarly, the revised mass $M'$ is given by, 
\begin{equation}
    M' = M \times (d' / d)^2\;.
\end{equation}
We note that these updates also lead to changes of other parameters, such as the virial parameter, which is given by,
\begin{equation}
    \alpha_{vir}' = \alpha_{vir} \times d / d'\;.
\end{equation}

Fig.\,\ref{xy} plots the space distribution of these 61 PGCCs in the $X$ -- $Y$ space with both the original kinematic distances and our updated distances, where the distance changes are indicated. According to the formula (3) and (4), we present the changes of these physical quantities in Fig.\,\ref{m_r}. From the histograms in the left panel of Fig.\,\ref{m_r}, we find that on the whole the radii and masses calculated with the new extinction distances tend to be smaller than those of \citet{Zhang2018}, by respectively $\sim$ 17.9 and 32.6 per\,cent. This leads us to conclude that to understand the local molecular structures, accurate distances are necessary. We use an MCMC algorithm to fit the mass-radius relation for both sets of distances. The new distances yield a slope of 1.95$\pm$0.09, comparable to 1.93$\pm$0.09 obtained with the kinematic distances. In spite of the significant changes, the slope is largely unaffected. The result is understandable, considering that a change in the distance measurement leads to changes in both mass and size estimates. Since the changes are correlated, the relation $M\sim R^2$ remains untouched.

In addition, we have applied simple Kolmogorov-Smirnov (K-S) tests to the newly derived physical properties (e.g. radius and masses) of the sample PGCCs and those from \citet{Zhang2018}. The distributions of the radius and masses obtained from the new extinction distances do not change significantly comparing to those estimated from the kinematic distances. Accurate extinction-based distance measurements are necessary for the estimation of robust physical parameters of the individual sources. However, as the sample PGCCs have relative large ranges of masses and radius, the distance error of a level of about 50 per cent will not significantly affect the overall sample
properties like the mass and radii distributions.

We have also revisited the relation between mass $M$\,($\rm M_{\odot}$) and virial parameter, $\alpha_{vir}$ = $M{_{vir}}$/{$M$} \citep{Bertoldi1992, Zhang2018}, the result of which is presented in the right panel of Fig.\,\ref{m_r}. The two slopes, $-$0.38$\pm$0.04 derived from the updated distances, and $-$0.39$\pm$0.04 derived from the kinematic distances, are consistent with each other. Again, the relation is largely unaffected, although the virial parameters of the individual cores do change significantly with the new distance estimates. On average, the new virial parameters of the cores deduced with our extinction distances are larger than those of \citet{Zhang2018} by $\sim$ 21.8 per\,cent. This again points to the importance of robust distance estimates when studying the gravitational boundedness of molecular clumps.

\section{Summary}

Based on results of 3D extinction mapping, we have calculated the extinction distances of 61 PGCCs in the second quadrant of the Milky Way analyzed previously by \citet{Zhang2018}. For this purpose, we use two catalogues published by \citet{Chen2019} and \citet{Green2019} that presented robust estimates of extinction and distance for huge numbers of stars. Distances of the clumps are determined by finding the positions of significant reddening jumps in the reddening profiles. The jumps are fitted with a simple reddening model using an MCMC algorithm. The typical uncertainties of distances thus derived are less than 8 per\,cent. For the majority of the PGCCs, we find significant differences between our newly derived distances and those kinematic ones from \citet{Zhang2018}.

Using newly estimated distances, we have updated distance-dependent physical parameters of the cores, including radius, mass and the virial parameter. Comparing to the results of \citet{Zhang2018}, the new radii, masses and virial parameters calculated using our new extinction distances tend to be smaller by $\sim$ 17.9, 32.6 and larger by 21.8 per\,cent, respectively.

\clearpage
\begin{table*}
 \begin{center}
 \centerline{\small {\bf Table~1.} Distances of 64 PGCCs in the second Galactic quadrant }

 \label{parameter}
 \begin{tabular}{|l|c|c|c|c|c|}
 \hline
 Name & R.A.(J2000) & DEC.(J2000)  & $d_{0}$ & $\delta E(B-V)$  & $d_{\rm Z18}$$^{a}$    \\
    & (hh:mm:ss) & (dd:mm:ss)  & (pc) & (mag) & (pc)\\
 \hline

G098.50-03.24 & 22:05:00.08 & +51:33:11.69 & $ 411 \pm 42 $ & 0.3 & $ 1590 \pm 250 $ \\
G108.85-00.80 & 22:58:51.53 & +58:57:27.09 & $ 2983 \pm 118 $ & 0.87 & $ 3210 \pm 380 $ \\
G110.65+09.65 & 22:28:00.22 & +69:01:48.10 & $ 351 \pm 35 $ & 0.6 & $ 820 \pm 110 $ \\
G112.52+08.38 & 22:52:47.62 & +68:49:28.31 & $ 363 \pm 13 $ & 0.89 & $ 780 \pm 100 $ \\
G112.60+08.53 & 22:52:54.76 & +68:59:53.90 & $ 346 \pm 11 $ & 0.86 & $ 780 \pm 100 $ \\
G115.92+09.46 & 23:24:04.62 & +71:08:08.69 & $ 258 \pm 26 $ & 0.59 & $ 730 \pm 80 $ \\
G116.08-02.38 & 23:56:41.79 & +59:45:13.19 & $ 371 \pm 12 $ & 0.87 & $ 720 \pm 90 $ \\
G116.12+08.98 & 23:28:14.03 & +70:45:12.38 & $ 265 \pm 15 $ & 0.56 & $ 730 \pm 80 $ \\
G120.16+03.09 & 00:24:26.01 & +65:49:27.59 & $ 1076 \pm 45 $ & 1.1 & $ 1280 \pm 650 $ \\
G120.67+02.66 & 00:29:41.95 & +65:26:39.99 & $ 1063 \pm 104 $ & 0.83 & $ 900 \pm 290 $ \\
G120.98+02.66 & 00:32:38.94 & +65:28:07.08 & $ 973 \pm 91 $ & 0.71 & $ 920 \pm 40 $ \\
G121.35+03.39 & 00:35:48.66 & +66:13:13.29 & $ 648 \pm 53 $ & 0.69 & $ 700 \pm 80 $ \\
G121.90-01.54 & 00:42:52.64 & +61:18:23.20 & $ 916 \pm 28 $ & 0.81 & $ 590 \pm 200 $ \\
G121.92-01.71 & 00:43:06.34 & +61:08:21.59 & $ 908 \pm 40 $ & 0.65 & $ 580 \pm 200 $ \\
G125.66-00.55 & 01:14:52.20 & +62:11:16.60 & $ 955 \pm 32 $ & 0.57 & $ 610 \pm 160 $ \\
G126.49-01.30 & 01:21:14.55 & +61:21:34.60 & $ 941 \pm 27 $ & 0.68 & $ 930 \pm 150 $ \\
G126.95-01.06 & 01:25:19.48 & +61:32:36.19 & $ 924 \pm 27 $ & 0.41 & $ 600 \pm 170 $ \\
G127.22-02.25 & 01:26:10.18 & +60:19:29.30 & $ 878 \pm 93 $ & 0.39 & $ 880 \pm 190 $ \\
G127.88+02.66 & 01:38:39.10 & +65:05:06.49 & $ 887 \pm 61 $ & 0.42 & $ 890 \pm 110 $ \\
G128.95-00.18 & 01:43:15.17 & +62:04:39.09 & $ 991 \pm 58 $ & 0.47 & $ 920 \pm 180 $ \\
G131.72+09.70 & 02:39:57.51 & +70:42:11.60 & $ 880 \pm 52 $ & 0.4 & $ 570 \pm 160 $ \\
G132.07+08.80 & 02:39:18.17 & +69:44:01.11 & $ 947 \pm 37 $ & 0.71 & $ 590 \pm 150 $ \\
G132.03+08.95 & 02:39:33.56 & +69:53:21.08 & $ 927 \pm 60 $ & 0.56 & $ 590 \pm 150 $ \\
G133.28+08.81 & 02:51:42.22 & +69:14:13.39 & $ 969 \pm 21 $ & 1.1 & $ 580 \pm 150 $ \\
G133.48+09.02 & 02:54:44.50 & +69:19:57.59 & $ 980 \pm 31 $ & 1.03 & $ 610 \pm 140 $ \\
G136.31-01.77 & 02:36:07.02 & +58:21:09.09 & $ 822 \pm 41 $ & 0.53 & $ 510 \pm 170 $ \\
G140.49+06.07 & 03:37:46.12 & +63:07:27.29 & $ 999 \pm 38 $ & 0.84 & $ 1230 \pm 490 $ \\
G140.77+05.00 & 03:34:18.18 & +62:05:35.89 & $ 997 \pm 61 $ & 0.67 & $ 560 \pm 140 $ \\
G142.49+07.48 & 03:59:13.56 & +62:58:52.40 & $ 930 \pm 35 $ & 0.44 & $ 550 \pm 140 $ \\
G142.62+07.29 & 03:59:00.66 & +62:45:12.60 & $ 942 \pm 33 $ & 0.46 & $ 540 \pm 140 $ \\
G144.84+00.76 & 03:40:20.80 & +56:16:28.09 & $ 1255 \pm 29 $ & 0.98 & $ 2200 \pm 280 $ \\
G146.11+07.80 & 04:23:14.52 & +60:44:31.20 & $ 655 \pm 47 $ & 0.56 & $ 520 \pm 140 $ \\
G146.71+02.05 & 03:56:37.16 & +56:07:23.10 & $ 243 \pm 16 $ & 0.79 & $ 470 \pm 150 $ \\
G147.01+03.39 & 04:04:41.36 & +56:56:16.79 & $ 218 \pm 25 $ & 0.53 & $ 500 \pm 140 $ \\
G148.00+00.09 & 03:54:48.04 & +53:47:19.89 & $ 1327 \pm 42 $ & 0.97 & $ 2150 \pm 280 $ \\
G148.24+00.41 & 03:57:26.18 & +53:52:36.30 & $ 1616 \pm 78 $ & 0.91 & $ 2170 \pm 290 $ \\
G149.23+03.07 & 04:14:48.52 & +55:12:03.29 & $ 306 \pm 23 $ & 1.4 & $ 470 \pm 150 $ \\
G149.41+03.37 & 04:17:09.06 & +55:17:39.39 & $ 271 \pm 20 $ & 1.16 & $ 470 \pm 150 $ \\
G149.52-01.23 & 03:56:52.61 & +51:48:01.70 & $ 1154 \pm 124 $ & 1.05 & $ 510 \pm 140 $ \\
G149.58+03.45 & 04:18:23.93 & +55:13:30.59 & $ 286 \pm 35 $ & 1.1 & $ 470 \pm 150 $ \\
G149.65+03.54 & 04:19:11.24 & +55:14:44.39 & $ 342 \pm 20 $ & 1.26 & $ 470 \pm 150 $ \\
G150.22+03.91 & 04:23:51.69 & +55:06:22.50 & $ 372 \pm 30 $ & 1.17 & $ 460 \pm 140 $ \\
G150.44+03.95 & 04:25:07.08 & +54:58:32.39 & $ 344 \pm 24 $ & 1.04 & $ 460 \pm 140 $ \\
G151.08+04.46 & 04:30:42.87 & +54:51:53.89 & $ 332 \pm 28 $ & 0.9 & $ 460 \pm 140 $ \\
G151.45+03.95 & 04:29:56.25 & +54:14:51.70 & $ 205 \pm 15 $ & 1.08 & $ 460 \pm 140 $ \\
G154.90+04.61 & 04:48:27.03 & +52:06:30.39 & $ 278 \pm 26 $ & 0.88 & $ 450 \pm 140 $ \\
G156.04+06.03 & 05:00:19.24 & +52:06:45.60 & $ 509 \pm 44 $ & 0.7 & $ 420 \pm 130 $ \\
G156.20+05.26 & 04:57:00.65 & +51:31:08.89 & $ 390 \pm 36 $ & 1.03 & $ 440 \pm 140 $ \\
G157.25-01.00 & 04:32:09.45 & +46:37:25.00 & $ 334 \pm 20 $ & 1.36 & $ 460 \pm 140 $ \\
G159.52+03.26 & 04:59:55.05 & +47:40:52.00 & $ 3685 \pm 505 $ & 1.04 & $ 1970 \pm 350 $ \\

\hline
 \end{tabular}
  \end{center}
  \end{table*}

\begin{table*}
 \begin{center}
 \centerline{\small {\bf Table~2} - continued }
 \label{parameter}
 \begin{tabular}{|l|c|c|c|c|c|}
 \hline
 Name & R.A.(J2000) & DEC.(J2000)  & $d_{0}$ & $\delta E(B-V)$  & $d_{\rm Z18}$$^{a}$       \\
    & (hh:mm:ss) & (dd:mm:ss)  & (pc) & (mag) & (pc)\\
 \hline
G162.79+01.34 & 05:02:42.87 & +43:55:05.70 & $ 477 \pm 35 $ & 0.87 & $ 450 \pm 140 $ \\
G169.14-01.13 & 05:12:20.07 & +37:20:57.09 & $ 3019 \pm 239 $ & 0.86 & $ 1870 \pm 230 $ \\
G171.03+02.66 & 05:33:35.43 & +37:56:42.69 & $ 1494 \pm 56 $ & 0.97 & $ 1820 \pm 220 $ \\
G171.34+02.59 & 05:34:06.95 & +37:38:47.30 & $ 1618 \pm 65 $ & 0.87 & $ 1780 \pm 200 $ \\
G172.85+02.27 & 05:36:51.80 & +36:11:58.29 & $ 1699 \pm 50 $ & 1.18 & $ 1710 \pm 130 $ \\
G175.20+01.28 & 05:38:55.10 & +33:41:05.89 & $-$ & 0.37 & $ 1690 \pm 120 $ \\
G175.53+01.34 & 05:39:59.24 & +33:26:08.80 & $-$ & 0.37 & $ 1690 \pm 120 $ \\
G176.17-02.10 & 05:27:55.18 & +31:01:34.99 & $ 1503 \pm 96 $ & 0.63 & $ 4960 \pm 440 $ \\
G176.35+01.92 & 05:44:23.17 & +33:02:58.99 & $-$  & 0.73 & $ 1700 \pm 130 $ \\
G176.94+04.63 & 05:57:00.77 & +33:55:16.30 & $ 1226 \pm 113 $ & 0.89 & $ 1780 \pm 220 $ \\
G177.09+02.85 & 05:50:02.12 & +32:53:35.90 & $ 4037 \pm 356 $ & 0.82 & $ 4970 \pm 570 $ \\
G177.14-01.21 & 05:33:52.82 & +30:42:36.29 & $ 1592 \pm 92 $ & 0.68 & $ 5000 \pm 480 $ \\
G177.86+01.04 & 05:44:35.76 & +31:17:57.40 & $ 1288 \pm 136 $ & 0.83 & $ 4990 \pm 640 $ \\
G178.28-00.61 & 05:39:03.83 & +30:04:05.90 & $ 1435 \pm 130 $ & 1.02 & $ 960 \pm 20 $ \\

\hline
 \end{tabular}
  \end{center}
\footnotesize{$^a$ Kinematic distances from \citet{Zhang2018}}\\
\end{table*}

\section*{Acknowledgements}

We thank the anonymous referee for the instructive comments. This work is partially supported by National Key R\&D Program of China No.~2019YFA0405503, National Natural Science Foundation of China grants No.~11803029, U1531244, 11833006 and U1731308, and Yunnan University grant No.~C176220100007.

\section*{Data availability}
The data underlying this article are available in the article and in its online supplementary material.

\bibliographystyle{mnras}
\bibliography{ms}

\begin{thebibliography}{}
\makeatletter
\relax
\def\mn@urlcharsother{\let\do\@makeother \do\$\do\&\do\#\do\^\do\_\do\%\do\~}
\def\mn@doi{\begingroup\mn@urlcharsother \@ifnextchar [ {\mn@doi@}
  {\mn@doi@[]}}
\def\mn@doi@[#1]#2{\def\@tempa{#1}\ifx\@tempa\@empty \href
  {http://dx.doi.org/#2} {doi:#2}\else \href {http://dx.doi.org/#2} {#1}\fi
  \endgroup}
\def\mn@eprint#1#2{\mn@eprint@#1:#2::\@nil}
\def\mn@eprint@arXiv#1{\href {http://arxiv.org/abs/#1} {{\tt arXiv:#1}}}
\def\mn@eprint@dblp#1{\href {http://dblp.uni-trier.de/rec/bibtex/#1.xml}
  {dblp:#1}}
\def\mn@eprint@#1:#2:#3:#4\@nil{\def\@tempa {#1}\def\@tempb {#2}\def\@tempc
  {#3}\ifx \@tempc \@empty \let \@tempc \@tempb \let \@tempb \@tempa \fi \ifx
  \@tempb \@empty \def\@tempb {arXiv}\fi \@ifundefined
  {mn@eprint@\@tempb}{\@tempb:\@tempc}{\expandafter \expandafter \csname
  mn@eprint@\@tempb\endcsname \expandafter{\@tempc}}}

\bibitem[\protect\citeauthoryear{{Anders} et~al.,}{{Anders}
  et~al.}{2019}]{Anders2019}
{Anders} F.,  et~al., 2019, \mn@doi [\aap] {10.1051/0004-6361/201935765}, \href
  {https://ui.adsabs.harvard.edu/abs/2019A&A...628A..94A} {628, A94}

\bibitem[\protect\citeauthoryear{{Bailer-Jones}, {Rybizki}, {Fouesneau},
  {Mantelet}  \& {Andrae}}{{Bailer-Jones} et~al.}{2018}]{Bailer2018}
{Bailer-Jones} C.~A.~L.,  {Rybizki} J.,  {Fouesneau} M.,  {Mantelet} G.,
  {Andrae} R.,  2018, \mn@doi [\aj] {10.3847/1538-3881/aacb21}, \href
  {https://ui.adsabs.harvard.edu/abs/2018AJ....156...58B} {156, 58}

\bibitem[\protect\citeauthoryear{{Bertoldi} \& {McKee}}{{Bertoldi} \&
  {McKee}}{1992}]{Bertoldi1992}
{Bertoldi} F.,  {McKee} C.~F.,  1992, \mn@doi [\apj] {10.1086/171638}, \href
  {https://ui.adsabs.harvard.edu/abs/1992ApJ...395..140B} {395, 140}

\bibitem[\protect\citeauthoryear{{Blitz} \& {Williams}}{{Blitz} \&
  {Williams}}{1999}]{Blitz1999}
{Blitz} L.,  {Williams} J.~P.,  1999, in {Lada} C.~J.,  {Kylafis} N.~D.,  eds,
  NATO Advanced Science Institutes (ASI) Series C Vol. 540, NATO Advanced
  Science Institutes (ASI) Series C. p.~3

\bibitem[\protect\citeauthoryear{{Chambers} et~al.,}{{Chambers}
  et~al.}{2016}]{Chambers2016}
{Chambers} K.~C.,  et~al., 2016, arXiv e-prints, \href
  {https://ui.adsabs.harvard.edu/abs/2016arXiv161205560C} {p. arXiv:1612.05560}

\bibitem[\protect\citeauthoryear{{Chen} et~al.,}{{Chen}
  et~al.}{2014}]{Chen2014}
{Chen} B.~Q.,  et~al., 2014, \mn@doi [\mnras] {10.1093/mnras/stu1192}, \href
  {https://ui.adsabs.harvard.edu/abs/2014MNRAS.443.1192C} {443, 1192}

\bibitem[\protect\citeauthoryear{{Chen} et~al.,}{{Chen}
  et~al.}{2017}]{Chen2017b}
{Chen} B.~Q.,  et~al., 2017, \mn@doi [\mnras] {10.1093/mnras/stx2287}, \href
  {https://ui.adsabs.harvard.edu/abs/2017MNRAS.472.3924C} {472, 3924}

\bibitem[\protect\citeauthoryear{{Chen} et~al.,}{{Chen}
  et~al.}{2019}]{Chen2019}
{Chen} B.~Q.,  et~al., 2019, \mn@doi [\mnras] {10.1093/mnras/sty3341}, \href
  {https://ui.adsabs.harvard.edu/abs/2019MNRAS.483.4277C} {483, 4277}

\bibitem[\protect\citeauthoryear{{Chen}, {Wang}, {Hou}, {Yang}, {Li}, {Zhao}
  \& {Jiang}}{{Chen} et~al.}{2020a}]{Chen2020b}
{Chen} B.,  {Wang} S.,  {Hou} L.,  {Yang} Y.,  {Li} Z.,  {Zhao} H.,   {Jiang}
  B.,  2020a, \mn@doi [\mnras] {10.1093/mnras/staa1827}, \href
  {https://ui.adsabs.harvard.edu/abs/2020MNRAS.tmp.1976C} {}

\bibitem[\protect\citeauthoryear{{Chen} et~al.,}{{Chen}
  et~al.}{2020b}]{Chen2020}
{Chen} B.~Q.,  et~al., 2020b, \mn@doi [\mnras] {10.1093/mnras/staa235}, \href
  {https://ui.adsabs.harvard.edu/abs/2020MNRAS.493..351C} {493, 351}

\bibitem[\protect\citeauthoryear{{Choi}, {Hachisuka}, {Reid}, {Xu},
  {Brunthaler}, {Menten}  \& {Dame}}{{Choi} et~al.}{2014}]{Choi2014}
{Choi} Y.~K.,  {Hachisuka} K.,  {Reid} M.~J.,  {Xu} Y.,  {Brunthaler} A.,
  {Menten} K.~M.,   {Dame} T.~M.,  2014, \mn@doi [\apj]
  {10.1088/0004-637X/790/2/99}, \href
  {https://ui.adsabs.harvard.edu/abs/2014ApJ...790...99C} {790, 99}

\bibitem[\protect\citeauthoryear{{Foreman-Mackey}, {Hogg}, {Lang}  \&
  {Goodman}}{{Foreman-Mackey} et~al.}{2013}]{Foreman2013}
{Foreman-Mackey} D.,  {Hogg} D.~W.,  {Lang} D.,   {Goodman} J.,  2013, \mn@doi
  [PASP] {10.1086/670067}, 125, 306

\bibitem[\protect\citeauthoryear{{Gaia Collaboration} et~al.,}{{Gaia
  Collaboration} et~al.}{2016}]{Gaia2016}
{Gaia Collaboration} et~al., 2016, \mn@doi [\aap]
  {10.1051/0004-6361/201629272}, \href
  {https://ui.adsabs.harvard.edu/abs/2016A&A...595A...1G} {595, A1}

\bibitem[\protect\citeauthoryear{{Gaia Collaboration} et~al.,}{{Gaia
  Collaboration} et~al.}{2018}]{Gaia2018a}
{Gaia Collaboration} et~al., 2018, \mn@doi [\aap]
  {10.1051/0004-6361/201833051}, \href
  {https://ui.adsabs.harvard.edu/abs/2018A&A...616A...1G} {616, A1}

\bibitem[\protect\citeauthoryear{{G{\'o}mez}}{{G{\'o}mez}}{2006}]{Gomez2006}
{G{\'o}mez} G.~C.,  2006, \mn@doi [\aj] {10.1086/508412}, \href
  {https://ui.adsabs.harvard.edu/abs/2006AJ....132.2376G} {132, 2376}

\bibitem[\protect\citeauthoryear{{Green} et~al.,}{{Green}
  et~al.}{2015}]{Green2015}
{Green} G.~M.,  et~al., 2015, \mn@doi [\apj] {10.1088/0004-637X/810/1/25},
  \href {https://ui.adsabs.harvard.edu/abs/2015ApJ...810...25G} {810, 25}

\bibitem[\protect\citeauthoryear{{Green} et~al.,}{{Green}
  et~al.}{2018}]{Green2018}
{Green} G.~M.,  et~al., 2018, \mn@doi [\mnras] {10.1093/mnras/sty1008}, \href
  {https://ui.adsabs.harvard.edu/abs/2018MNRAS.478..651G} {478, 651}

\bibitem[\protect\citeauthoryear{{Green}, {Schlafly}, {Zucker}, {Speagle}  \&
  {Finkbeiner}}{{Green} et~al.}{2019}]{Green2019}
{Green} G.~M.,  {Schlafly} E.,  {Zucker} C.,  {Speagle} J.~S.,   {Finkbeiner}
  D.,  2019, \mn@doi [\apj] {10.3847/1538-4357/ab5362}, \href
  {https://ui.adsabs.harvard.edu/abs/2019ApJ...887...93G} {887, 93}

\bibitem[\protect\citeauthoryear{{Guo} et~al.,}{{Guo} et~al.}{2020}]{Guo2020}
{Guo} H.~L.,  et~al., 2020, arXiv e-prints, \href
  {https://ui.adsabs.harvard.edu/abs/2020arXiv201014092G} {p. arXiv:2010.14092}

\bibitem[\protect\citeauthoryear{{Hachisuka}, {Choi}, {Reid}, {Brunthaler},
  {Menten}, {Sanna}  \& {Dame}}{{Hachisuka} et~al.}{2015}]{Hachisuka2015}
{Hachisuka} K.,  {Choi} Y.~K.,  {Reid} M.~J.,  {Brunthaler} A.,  {Menten}
  K.~M.,  {Sanna} A.,   {Dame} T.~M.,  2015, \mn@doi [\apj]
  {10.1088/0004-637X/800/1/2}, \href
  {https://ui.adsabs.harvard.edu/abs/2015ApJ...800....2H} {800, 2}

\bibitem[\protect\citeauthoryear{{Holland} et~al.,}{{Holland}
  et~al.}{2013}]{Holland2013}
{Holland} W.~S.,  et~al., 2013, \mn@doi [\mnras] {10.1093/mnras/sts612}, \href
  {https://ui.adsabs.harvard.edu/abs/2013MNRAS.430.2513H} {430, 2513}

\bibitem[\protect\citeauthoryear{{Huang} et~al.,}{{Huang}
  et~al.}{2016}]{Huang2016}
{Huang} Y.,  et~al., 2016, \mn@doi [\mnras] {10.1093/mnras/stw2096}, \href
  {https://ui.adsabs.harvard.edu/abs/2016MNRAS.463.2623H} {463, 2623}

\bibitem[\protect\citeauthoryear{{Kolpak}, {Jackson}, {Bania}, {Clemens}  \&
  {Dickey}}{{Kolpak} et~al.}{2003}]{Kolpak2003}
{Kolpak} M.~A.,  {Jackson} J.~M.,  {Bania} T.~M.,  {Clemens} D.~P.,   {Dickey}
  J.~M.,  2003, \mn@doi [\apj] {10.1086/344752}, \href
  {https://ui.adsabs.harvard.edu/abs/2003ApJ...582..756K} {582, 756}

\bibitem[\protect\citeauthoryear{{Lallement}, {Babusiaux}, {Vergely}, {Katz},
  {Arenou}, {Valette}, {Hottier}  \& {Capitanio}}{{Lallement}
  et~al.}{2019}]{Lallement2019}
{Lallement} R.,  {Babusiaux} C.,  {Vergely} J.~L.,  {Katz} D.,  {Arenou} F.,
  {Valette} B.,  {Hottier} C.,   {Capitanio} L.,  2019, \mn@doi [\aap]
  {10.1051/0004-6361/201834695}, \href
  {https://ui.adsabs.harvard.edu/abs/2019A&A...625A.135L} {625, A135}

\bibitem[\protect\citeauthoryear{{Lindegren} et~al.,}{{Lindegren}
  et~al.}{2018}]{Lindegren2018}
{Lindegren} L.,  et~al., 2018, \mn@doi [\aap] {10.1051/0004-6361/201832727},
  \href {https://ui.adsabs.harvard.edu/abs/2018A&A...616A...2L} {616, A2}

\bibitem[\protect\citeauthoryear{{Liszt} \& {Burton}}{{Liszt} \&
  {Burton}}{1981}]{Liszt1981}
{Liszt} H.~S.,  {Burton} W.~B.,  1981, \mn@doi [\apj] {10.1086/158646}, \href
  {https://ui.adsabs.harvard.edu/abs/1981ApJ...243..778L} {243, 778}

\bibitem[\protect\citeauthoryear{{Pestalozzi}, {Minier}  \&
  {Booth}}{{Pestalozzi} et~al.}{2005}]{Pestalozzi2005}
{Pestalozzi} M.~R.,  {Minier} V.,   {Booth} R.~S.,  2005, \mn@doi [\aap]
  {10.1051/0004-6361:20035855}, \href
  {https://ui.adsabs.harvard.edu/abs/2005A&A...432..737P} {432, 737}

\bibitem[\protect\citeauthoryear{{Planck Collaboration} et~al.,}{{Planck
  Collaboration} et~al.}{2011a}]{Planck2011a}
{Planck Collaboration} et~al., 2011a, \mn@doi [\aap]
  {10.1051/0004-6361/201116464}, \href
  {https://ui.adsabs.harvard.edu/abs/2011A&A...536A...1P} {536, A1}

\bibitem[\protect\citeauthoryear{{Planck Collaboration} et~al.,}{{Planck
  Collaboration} et~al.}{2011b}]{Planck2011b}
{Planck Collaboration} et~al., 2011b, \mn@doi [\aap]
  {10.1051/0004-6361/201116474}, \href
  {https://ui.adsabs.harvard.edu/abs/2011A&A...536A...7P} {536, A7}

\bibitem[\protect\citeauthoryear{{Planck Collaboration} et~al.,}{{Planck
  Collaboration} et~al.}{2011c}]{Planck2011c}
{Planck Collaboration} et~al., 2011c, \mn@doi [\aap]
  {10.1051/0004-6361/201116481}, \href
  {https://ui.adsabs.harvard.edu/abs/2011A&A...536A..22P} {536, A22}

\bibitem[\protect\citeauthoryear{{Planck Collaboration} et~al.,}{{Planck
  Collaboration} et~al.}{2011d}]{Planck2011d}
{Planck Collaboration} et~al., 2011d, \mn@doi [\aap]
  {10.1051/0004-6361/201116472}, \href
  {https://ui.adsabs.harvard.edu/abs/2011A&A...536A..23P} {536, A23}

\bibitem[\protect\citeauthoryear{{Reid} et~al.,}{{Reid}
  et~al.}{2009}]{Reid2009}
{Reid} M.~J.,  et~al., 2009, \mn@doi [\apj] {10.1088/0004-637X/700/1/137},
  \href {https://ui.adsabs.harvard.edu/abs/2009ApJ...700..137R} {700, 137}

\bibitem[\protect\citeauthoryear{{Reid} et~al.,}{{Reid}
  et~al.}{2014}]{Reid2014}
{Reid} M.~J.,  et~al., 2014, \mn@doi [\apj] {10.1088/0004-637X/783/2/130},
  \href {https://ui.adsabs.harvard.edu/abs/2014ApJ...783..130R} {783, 130}

\bibitem[\protect\citeauthoryear{{Reid}, {Dame}, {Menten}  \&
  {Brunthaler}}{{Reid} et~al.}{2016}]{Reid2016}
{Reid} M.~J.,  {Dame} T.~M.,  {Menten} K.~M.,   {Brunthaler} A.,  2016, \mn@doi
  [\apj] {10.3847/0004-637X/823/2/77}, \href
  {https://ui.adsabs.harvard.edu/abs/2016ApJ...823...77R} {823, 77}

\bibitem[\protect\citeauthoryear{{Schlafly} et~al.,}{{Schlafly}
  et~al.}{2014}]{Schlafly2014}
{Schlafly} E.~F.,  et~al., 2014, \mn@doi [\apj] {10.1088/0004-637X/786/1/29},
  \href {https://ui.adsabs.harvard.edu/abs/2014ApJ...786...29S} {786, 29}

\bibitem[\protect\citeauthoryear{{Shan}, {Zhu}, {Tian}, {Zhang}, {Yang}  \&
  {Zhang}}{{Shan} et~al.}{2019}]{Shan2019}
{Shan} S.-S.,  {Zhu} H.,  {Tian} W.-W.,  {Zhang} H.-Y.,  {Yang} A.-Y.,
  {Zhang} M.-F.,  2019, \mn@doi [Research in Astronomy and Astrophysics]
  {10.1088/1674-4527/19/7/92}, \href
  {https://ui.adsabs.harvard.edu/abs/2019RAA....19...92S} {19, 092}

\bibitem[\protect\citeauthoryear{{Skrutskie} et~al.,}{{Skrutskie}
  et~al.}{2006}]{Skrutskie2006}
{Skrutskie} M.~F.,  et~al., 2006, \mn@doi [\aj] {10.1086/498708}, \href
  {https://ui.adsabs.harvard.edu/abs/2006AJ....131.1163S} {131, 1163}

\bibitem[\protect\citeauthoryear{{Wall} \& {Jenkins}}{{Wall} \&
  {Jenkins}}{2003}]{Jenkins2003}
{Wall} J.~V.,  {Jenkins} C.~R.,  2003, {Practical Statistics for Astronomers}.
 Vol. 3

\bibitem[\protect\citeauthoryear{{Wang}, {Zhang}, {Jiang}, {Zhao}, {Chen},
  {Chen}, {Gao}  \& {Liu}}{{Wang} et~al.}{2020}]{Wang2020}
{Wang} S.,  {Zhang} C.,  {Jiang} B.,  {Zhao} H.,  {Chen} B.,  {Chen} X.,  {Gao}
  J.,   {Liu} J.,  2020, \mn@doi [\aap] {10.1051/0004-6361/201936868}, \href
  {https://ui.adsabs.harvard.edu/abs/2020A&A...639A..72W} {639, A72}

\bibitem[\protect\citeauthoryear{{Wright} et~al.,}{{Wright}
  et~al.}{2010}]{Wright2010}
{Wright} E.~L.,  et~al., 2010, \mn@doi [\aj] {10.1088/0004-6256/140/6/1868},
  \href {https://ui.adsabs.harvard.edu/abs/2010AJ....140.1868W} {140, 1868}

\bibitem[\protect\citeauthoryear{{Wu}, {Liu}, {Meng}, {Li}, {Qin}  \&
  {Ju}}{{Wu} et~al.}{2012}]{Wu2012}
{Wu} Y.,  {Liu} T.,  {Meng} F.,  {Li} D.,  {Qin} S.-L.,   {Ju} B.-G.,  2012,
  \mn@doi [\apj] {10.1088/0004-637X/756/1/76}, \href
  {https://ui.adsabs.harvard.edu/abs/2012ApJ...756...76W} {756, 76}

\bibitem[\protect\citeauthoryear{{Wu} et~al.,}{{Wu} et~al.}{2014}]{Wu2014}
{Wu} Y.~W.,  et~al., 2014, \mn@doi [\aap] {10.1051/0004-6361/201322765}, \href
  {https://ui.adsabs.harvard.edu/abs/2014A&A...566A..17W} {566, A17}

\bibitem[\protect\citeauthoryear{{Xu} et~al.,}{{Xu} et~al.}{2013}]{Xu2013}
{Xu} Y.,  et~al., 2013, \mn@doi [\apj] {10.1088/0004-637X/769/1/15}, \href
  {https://ui.adsabs.harvard.edu/abs/2013ApJ...769...15X} {769, 15}

\bibitem[\protect\citeauthoryear{{Xu} et~al.,}{{Xu} et~al.}{2016}]{Xu2016}
{Xu} Y.,  et~al., 2016, \mn@doi [Science Advances] {10.1126/sciadv.1600878},
  \href {https://ui.adsabs.harvard.edu/abs/2016SciA....2E0878X} {2, e1600878}

\bibitem[\protect\citeauthoryear{{Xue}, {Rix}, {Ma}, {Morrison}, {Bovy},
  {Sesar}  \& {Janesh}}{{Xue} et~al.}{2015}]{Xue2015}
{Xue} X.-X.,  {Rix} H.-W.,  {Ma} Z.,  {Morrison} H.,  {Bovy} J.,  {Sesar} B.,
  {Janesh} W.,  2015, \mn@doi [\apj] {10.1088/0004-637X/809/2/144}, \href
  {https://ui.adsabs.harvard.edu/abs/2015ApJ...809..144X} {809, 144}

\bibitem[\protect\citeauthoryear{{Yu}, {Chen}, {Jiang}  \& {Zijlstra}}{{Yu}
  et~al.}{2019}]{Yu2019}
{Yu} B.,  {Chen} B.~Q.,  {Jiang} B.~W.,   {Zijlstra} A.,  2019, \mn@doi
  [\mnras] {10.1093/mnras/stz1940}, \href
  {https://ui.adsabs.harvard.edu/abs/2019MNRAS.488.3129Y} {488, 3129}

\bibitem[\protect\citeauthoryear{{Zhang}, {Reid}, {Menten}, {Zheng},
  {Brunthaler}, {Dame}  \& {Xu}}{{Zhang} et~al.}{2013}]{Zhang2013}
{Zhang} B.,  {Reid} M.~J.,  {Menten} K.~M.,  {Zheng} X.~W.,  {Brunthaler} A.,
  {Dame} T.~M.,   {Xu} Y.,  2013, \mn@doi [\apj] {10.1088/0004-637X/775/1/79},
  \href {https://ui.adsabs.harvard.edu/abs/2013ApJ...775...79Z} {775, 79}

\bibitem[\protect\citeauthoryear{{Zhang}, {Wu}, {Liu}  \& {Meng}}{{Zhang}
  et~al.}{2016}]{Zhang2016}
{Zhang} T.,  {Wu} Y.,  {Liu} T.,   {Meng} F.,  2016, \mn@doi [\apjs]
  {10.3847/0067-0049/224/2/43}, \href
  {https://ui.adsabs.harvard.edu/abs/2016ApJS..224...43Z} {224, 43}

\bibitem[\protect\citeauthoryear{{Zhang} et~al.,}{{Zhang}
  et~al.}{2018}]{Zhang2018}
{Zhang} C.-P.,  et~al., 2018, \mn@doi [\apjs] {10.3847/1538-4365/aac513}, \href
  {https://ui.adsabs.harvard.edu/abs/2018ApJS..236...49Z} {236, 49}

\bibitem[\protect\citeauthoryear{{Zhao}, {Jiang}, {Gao}, {Li}  \& {Sun}}{{Zhao}
  et~al.}{2018}]{Zhao2018}
{Zhao} H.,  {Jiang} B.,  {Gao} S.,  {Li} J.,   {Sun} M.,  2018, \mn@doi [\apj]
  {10.3847/1538-4357/aaacd0}, \href
  {https://ui.adsabs.harvard.edu/abs/2018ApJ...855...12Z} {855, 12}

\bibitem[\protect\citeauthoryear{{Zhao}, {Jiang}, {Li}, {Chen}, {Yu}  \&
  {Wang}}{{Zhao} et~al.}{2020}]{Zhao2020}
{Zhao} H.,  {Jiang} B.,  {Li} J.,  {Chen} B.,  {Yu} B.,   {Wang} Y.,  2020,
  \mn@doi [\apj] {10.3847/1538-4357/ab75ef}, \href
  {https://ui.adsabs.harvard.edu/abs/2020ApJ...891..137Z} {891, 137}

\makeatother
\end{thebibliography}

\end{document}